\documentclass[a4paper,fleqn]{cas-sc}

\usepackage[numbers]{natbib}

\def\tsc#1{\csdef{#1}{\textsc{\lowercase{#1}}\xspace}}
\tsc{WGM}
\tsc{QE}
\tsc{EP}
\tsc{PMS}
\tsc{BEC}
\tsc{DE}
\usepackage{graphicx}
\usepackage{amsthm}
 \usepackage{booktabs}
 \usepackage{siunitx}
 \usepackage{float}
 \restylefloat{figure}
 \usepackage{multirow}
 \usepackage{caption}
 \usepackage{bm}
 \usepackage[table]{xcolor}
  
\begin{document}
\let\WriteBookmarks\relax
\def\floatpagepagefraction{1}
\def\textpagefraction{.001}

\shorttitle{Operator-Aligned PEB Reconstruction}

\shortauthors{Z.Tao et~al.}

\title{Operator-Consistent Physics-Informed Learning for Wafer Thermal Reconstruction in Lithography}                      
\tnotemark[1,2]


%
\author[1]{Ze Tao}[orcid=0009-0004-0202-3641]
\credit{Calculation, data analyzing and manuscript writing}
\author[1]{Fujun Liu}[orcid=0000-0002-8573-450X]
\credit{Review and Editing}
\cormark[1]
\ead{fjliu@cust.edu.cn}
\cortext[cor]{Corresponding author}
\author[1]{Yuxi Jin}[orcid=0009-0003-9926-8866]
\credit{Data analyzing}
\author[1]{Ke Xu}[orcid=0009-0003-7880-0235]
\credit{Data analyzing}
\author[1]{Minghui Sun}[orcid=0009-0004-9198-2376]
\credit{Data analyzing}
\author[1]{Xiangsheng Hu}[orcid=0009-0001-6432-2733]
\credit{Data analyzing}
\author[1]{Qi Cao}[orcid=0009-0004-5788-1328]
\credit{Data analyzing}
\author[1]{Haoran Xu}[orcid=0009-0008-1654-7648]
\credit{Data analyzing}
\affiliation[1]{organization={Nanophotonics and Biophotonics Key Laboratory of Jilin Province, School of Physics, Changchun University of Science and Technology},
                city={Changchun},
                postcode={130022},
                country={P.R. China}}
\author[2]{Hanxuan Wang}[orcid=0000-0003-1830-5913]
\credit{Calculation}
\affiliation[2]{organization={Faculty of Chinese Medicine, Macau University of Science and 
Technology},
                city={Macau},
                postcode={999078},
                country={P.R. China}}
\begin{abstract}
Thermal field reconstruction in post-exposure bake (PEB) is critical for advanced lithography, yet current physics-informed neural networks (PINNs) suffer from inconsistent accuracy due to a misalignment between geometric coordinates, physical fields, and differential operators. To resolve this, we introduce a novel architecture that unifies these elements on a single computation graph by integrating LSTM-gated mechanisms within a Liquid Neural Network (LNN) backbone. This specific combination of gated liquid layers is necessary to dynamically regulate the network's spectral behavior and enforce operator-level consistency, which ensures stable training and high-fidelity predictions. Applied to a 2D PEB scenario with internal heat generation and convective boundaries, our model formulates residuals via differential forms and a composite loss functional. The results demonstrate rapid convergence, uniformly low errors, strong agreement with FEM benchmarks, and stable training without late-stage oscillations, outperforming existing baselines in accuracy and robustness. Our framework thus establishes a reliable foundation for high-fidelity thermal modeling and offers a transferable strategy for operator-consistent neural surrogates in other physical domains.
\end{abstract}


\begin{highlights}
    \item A geometry-aligned PINN consistently enforces constitutive laws and boundary conditions.
    \item Physics residuals from differential forms ensure unified trace and operator-level fidelity.
    \item The model achieves domain-wide low error and stable, oscillation-free convergence.
\end{highlights}
\begin{graphicalabstract}
\includegraphics[width=\textwidth]{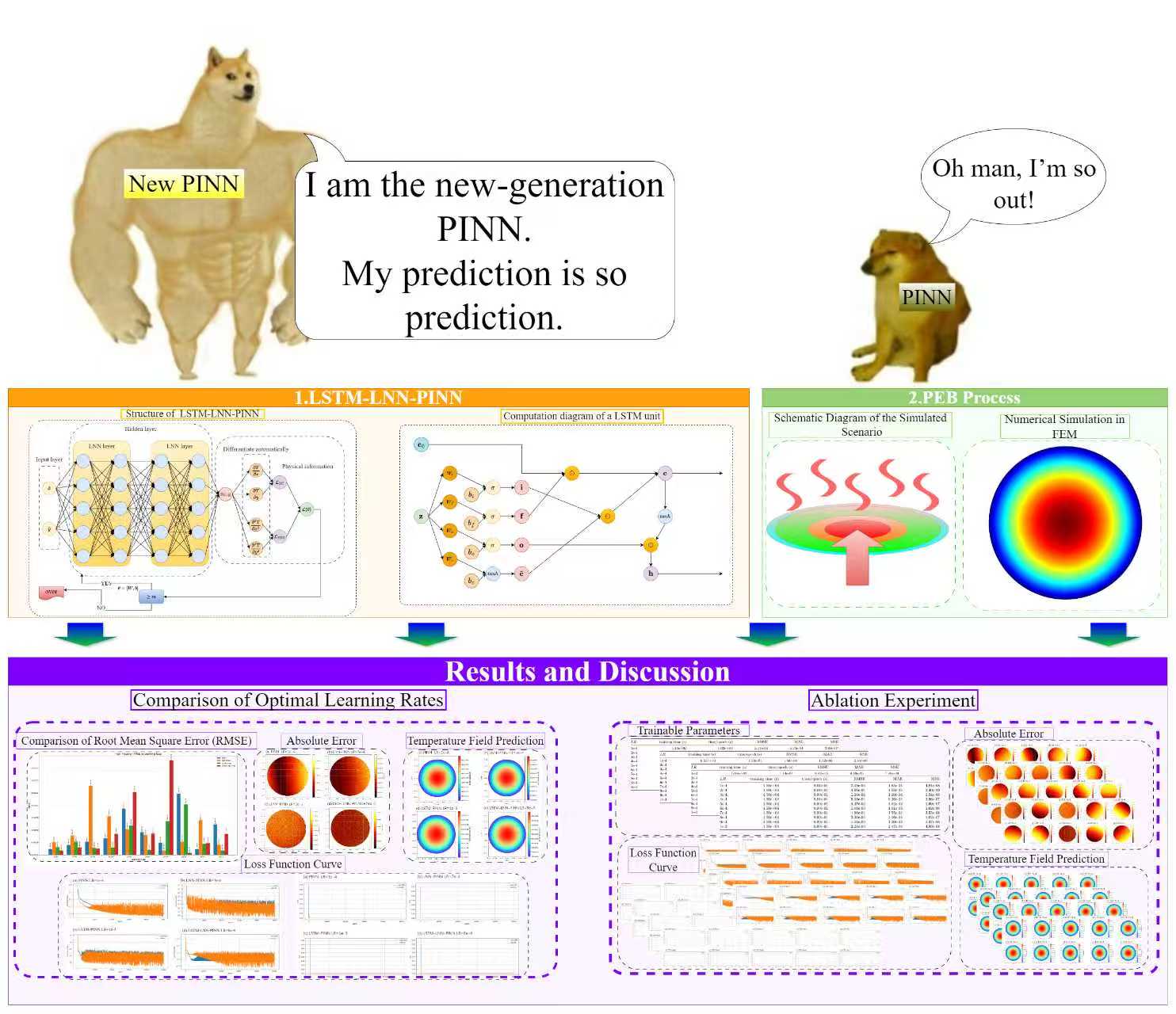}
\end{graphicalabstract}


\maketitle
\section{Introduction}\label{sec1}
Precise thermal management during post-exposure bake (PEB) is a critical bottleneck in advanced semiconductor manufacturing. Failure to control non-uniform temperature distributions directly leads to critical dimension variations, line-edge roughness, pattern transfer errors, and ultimately, catastrophic yield loss and reduced device reliability\cite{muneeshwaran2020thermal,jung2012deprotection,jiang2019chemical}. Precise thermal management during this phase impacts the fidelity of pattern transfer, yield, and device reliability in advanced integrated circuits \cite{ngo2012situ}. As wafer size and power density continue to scale, conventional heat dissipation strategies face increasing limitations in capturing non-uniform temperature distributions and geometry-sensitive flux patterns \cite{archenti2024integrated}. These limitations are particularly evident when high-resolution reconstructions are required across large-scale substrates with spatially heterogeneous boundary conditions \cite{liao2023hybrid,gong2021physics,li2021physics,chen2024accelerating}. Consequently, developing a thermal reconstruction framework that incorporates physical constraints while adapting to the complex geometry–port coupling inherent to wafer-scale conduction is essential for reliable process control and accurate inverse design \cite{romano2022inverse,erturk2023inverse,venuthurumilli2021inverse,nakagawa2023level}. 

Operator Alignment provides the central organizing principle\cite{lu2021learning}. Prior physics-informed networks decouple coordinates, fields, and differential operators across disjoint representations, which prevents a single computation graph from shaping the local spectrum and creates conflict between interior equations and boundary traces\cite{wang2021understanding,krishnapriyan2021characterizing,wang2021understanding}. The approach aligns the coordinate→field→operator chain within a typed computation graph by modeling the predictor $\Phi_\theta: \Omega \to H^1(\Omega)$\cite{kharazmi2021hp}. A single-step gated liquid residual layer implements input-conditioned diagonal preconditioning, constrains the spectral radius of hidden-to-hidden maps, preserves covariance, and stabilizes operator-aligned evaluations of second-order and trace operators\cite{hasani2021liquid,hasani2022closed,gouk2021regularisation}. A variational objective written in differential-form residuals attains energy equivalence with the weak formulation, thereby synchronizing the decay of interior and boundary losses. Under this alignment, the network Jacobian and Hessian align structurally with the physical operator, convergence stays non-oscillatory, and errors localize to geometry- and port-dominated regions. On post-exposure-bake reconstructions, the aligned model achieves lower global error and stronger robustness to learning-rate perturbations than misaligned baselines.

Recent developments in hybrid neural architectures offer a promising path toward resolving these limitations \cite{cuomo2022scientific,klapa2024machine}. Operator-aligned designs that incorporate gated dynamics, for instance, can regulate spectral behavior, stabilize higher-order derivatives, and preserve the structure of physical laws across computation graphs. By embedding single-step gating (acting as an input-conditioned diagonal preconditioner) within a liquid neural backbone, it becomes possible to dynamically control local conditioning while maintaining a unified pipeline from coordinates to fields to operators. This configuration not only retains the benefits of physical supervision but also adapts the predictor’s response to the intrinsic structure of the solution space, thereby establishing a consistent representation across layers and scales that is well-suited for high-fidelity thermal field reconstructions under process-driven constraints. We deploy this hybrid architecture on the PEB heat-flux reconstruction task and achieve high-fidelity results across all evaluation metrics. The predicted temperature field closely matches the benchmark solution, with uniformly low absolute errors maintained throughout the entire domain. Furthermore, residual losses in both governing equations and boundary conditions exhibit synchronized decay and settle into stable plateaus without late-stage fluctuations. These outcomes confirm that the model maintains operator-level consistency and geometric coherence during training, establishing a reliable foundation for wafer-scale thermal reconstruction under process-aligned constraints and marking a substantial advancement in physics-informed heat transfer modeling.

Unlike standard models, our architecture introduces three specific innovations: (1) It uses pointwise gating to create an input-dependent preconditioner, allowing the network to adapt its local spectral properties. (2) It decouples the LNN gates (f, i, o) to explicitly control 'contraction-step size-readout,' which provably stabilizes the evaluation of second-order operators. (3) It removes time unrolling, retaining the operator-aligned conditioning benefits of LSTM gating without the computational overhead of sequence modeling. We show this architecture is not just more accurate (achieving an RMSE of $3.83 \times 10^{-5}$ ) but uniquely achieves stable, oscillation-free convergence, solving a key failure mode of other PINN-based approaches.

\section{Operator-Aligned Network under Geometry–Port Consistency}

\begin{figure*}[p]
	\centering
	\includegraphics[width=0.8\textwidth]{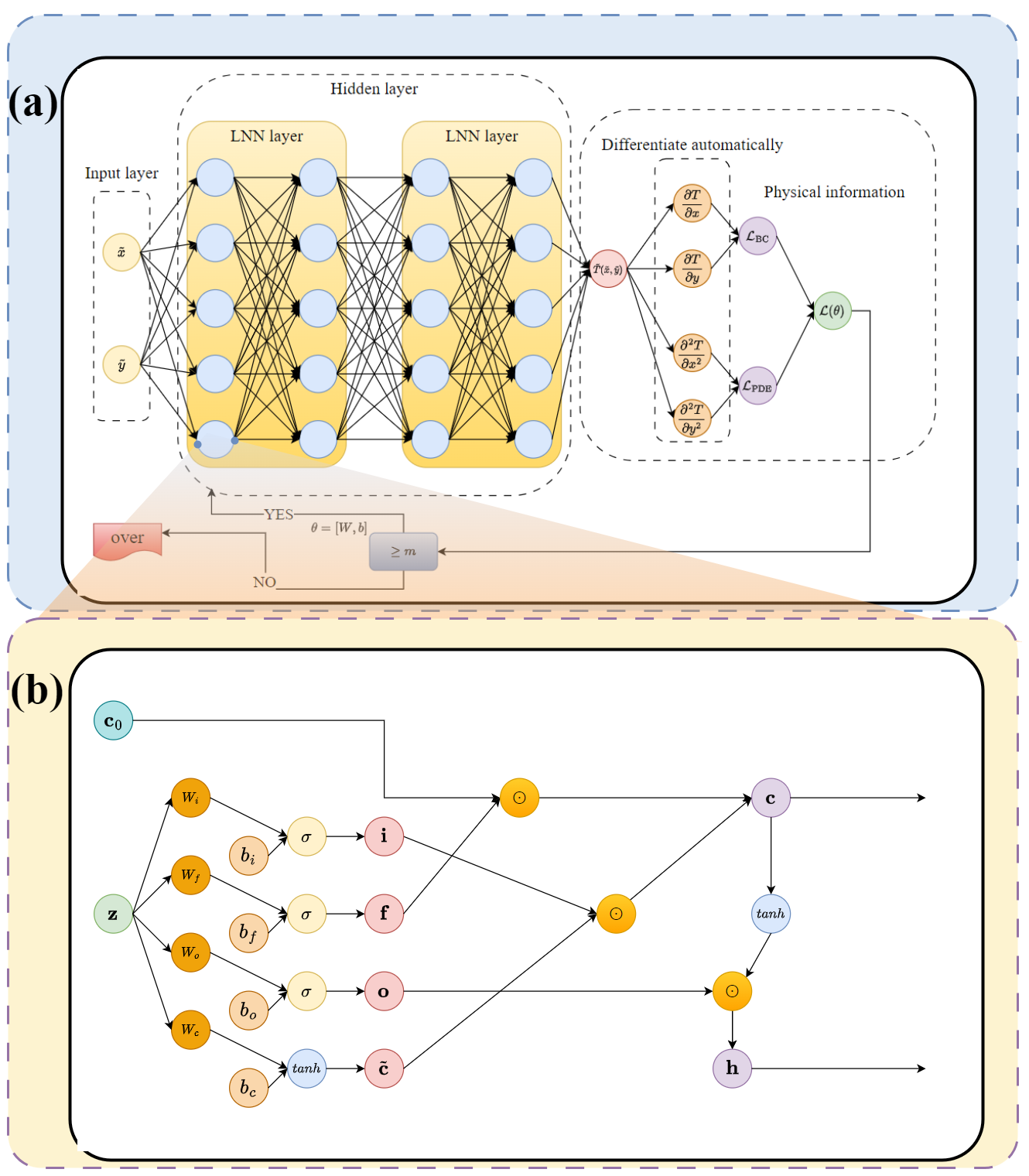}
	\caption{We present an LSTM-gated LNN-PINN architecture that maps spatial coordinates to temperature via a shared trunk and branches into physics heads for PDE, boundary/port, and source residuals. Residual channels aggregate into a single energy-consistent loss, stabilizing training and delivering high-fidelity predictions with accurate boundary traces. \textbf{(a)} Structure of the Physics-Informed Neural Network (LSTM-LNN-PINN) incorporating Gated Liquid Neural Networks. The original Multi-Layer Perceptron (MLP) is replaced by Liquid Neural Network (LNN) modules with Long Short-Term Memory (LSTM) gating mechanisms, forming a two-layer gated LNN network with 64 neurons per layer. This design dynamically adjusts information flow and feature extraction capabilities, enhancing modeling performance for spatiotemporal nonlinear behaviors in complex PEB process heat transfer fields while preserving physical constraints. The figure illustrates the interconnections and data flow among submodules, reflecting the deep integration of physical knowledge with neural network innovations. \textbf{(b)} Computation diagram of a Long Short-Term Memory (LSTM) unit, showing the input gate, forget gate, output gate, and candidate cell state. The figure illustrates affine transformations of weights and biases, nonlinear activation mappings, element-wise gate multiplications, weighted updates of previous cell state \(\mathbf{c}_{0}\), generation of current cell state \(\mathbf{c}\), and final hidden state \(\mathbf{h}\) modulated by \(\tanh\) and output gate. This demonstrates the memory retention and selective forgetting mechanisms inherent to Long Short-Term Memory (LSTM) units for sequence modeling in PEB process heat transfer fields while gates act as input-conditioned diagonal preconditioners along operator-aligned axes.
   }
	\label{F1}
\end{figure*}

\textbf{Key Architectural Innovations.} Operator alignment governs every architectural choice in this section, linking representation, dynamics, and the variational objective on a single computation graph. Several key distinctions differentiate our approach from existing architectures. Compared to static MLP–PINN, we place pointwise gating $(i,f,o)$ in every layer to induce an input-dependent diagonal preconditioner $D_{i,f,o}(x)$, ensuring the linearization satisfies $J(x)=D_{i,f,o}(x)W+S(x)$ with $S$ bounded and skew-symmetric. This allows inputs to adaptively reshape local spectra and control the amplification factor of $\nabla^2 T_\theta$, a property not available in the layerwise Jacobian of standard MLPs. Relative to LNN–PINN with leaky single gates, we decouple "leak/injection/readout" into three distinct gates $(f,i,o)$, emphasizing "contraction–step size–readout" in the main text. A small-gain argument pushes the spectral radius of the hidden-to-hidden map into a controllable interval $\rho<1$, thereby stabilizing second-order operator evaluation. In contrast to LSTM–PINN with explicit sequence axes, we retain the operator-aligned conditioning benefits of gating while removing time unrolling: single-step gating achieves first-order accuracy equivalent to a residual block with input-dependent diagonal preconditioning, maintaining computational complexity comparable to a same-width fully connected layer while avoiding extra backpropagation depth.
As illustrated in Fig.~\ref{F1} (a), this architecture replaces conventional MLPs with a dynamically evolving network that incorporates state differential equations and adaptive neuron coupling to enhance the joint modeling of local responses and global physical patterns in operator-aligned directions. The predictor is treated as a typed operator map $\Phi_\theta:\Omega\to H^1(\Omega)$, enabling the computation graph to simultaneously yield $\nabla T_\theta$ and $\nabla^2 T_\theta$ through automatic differentiation, thereby directly evaluating interior and boundary differential quantities along with required trace mappings.

The second derivatives of the network map are controlled through the Supplementary Material (SM) "sum–product" inequality (Eq. (S89)), which couples layerwise second-order constant \(H_\ell\) with first-order constant \(L_j\)
\begin{equation}
    \big\|D^{2}\Phi_{\theta}\big\|_{\mathrm{op}}
\;\le\;
\sum_{\ell=1}^{m}
\left(
  H_{\ell}\;\prod_{j=\ell+1}^{m} L_{j}^{\,2}\;\prod_{j=1}^{\ell-1} L_{j}
\right),
\quad L_j:=\sup_{z}\|DF_j(z)\|_{\mathrm{op}},\ \ H_j:=\sup_{z}\|D^2F_j(z)\|_{\mathrm{op}}.
\end{equation}
A tighter bound for a single gated layer is provided by Eq.~(S88)
\begin{equation}
\|D^{2}\Phi_{\theta}\|_{\mathrm{op}}\;\le\;\beta_g\,\|W\|^{2}\,\|W^{(\mathrm{out})}\|^{2},
\end{equation}
with transfer to physical second-order operators following from Eq.~(S41), yielding
\(\ \|\nabla^{2}T_{\theta}\|_{\mathrm{op}} \le C_{\mathrm{geom}}\|D^{2}\Phi_{\theta}\|_{\mathrm{op}}\). The network adopts a "single–gating, single–readout" feedforward execution without time unrolling, where gate variables $(i,f,o)$ modify only the local condition number and spectral shape of $\Phi\theta$ without altering the covariant relations between geometry and operators. Proofs of "preconditioned residual equivalence" and "product–type upper bound for the second derivative" are provided in the Sec. I of the Supplementary Material (SM), while the main text maintains focus on alignment between function–space types and operator channels. Consequently, gating modifies only the local condition number in operator-aligned directions and leaves the geometry–port–operator pipeline covariant by construction.

Spatial coordinates $\mathbf{x} = (x, y) \in \Omega \subset \mathbb{R}^{2}$ serve as network inputs, projected to the first LNN layer containing $d = 64$ liquid neurons. Each neuron receives combined current input and historical states, updates its internal state, and propagates outputs to subsequent layers. The internal state $\mathbf{h}{i}(t) \in \mathbb{R}$ of each liquid neuron follows a continuous-time dynamic system
\begin{equation}
\frac{d \mathbf{h}_{i}(t)}{d t} = -\lambda_{i} \mathbf{h}_{i}(t) + \phi_{i}\left(\mathbf{x}, \mathbf{h}_{i}(t) ; \theta_{i}\right)
\end{equation}
where $\lambda_{i} > 0$ denotes the leak coefficient and $\phi_{i}$ represents a nonlinear response function determined by neuron coupling, input projection, and bias. For numerical implementation, explicit Euler integration provides one-step approximation
\begin{equation}
\mathbf{h}_{i} \approx \alpha_{i} \mathbf{h}_{i}^{(0)} + (1 - \alpha_{i}) \phi_{i}\left(\mathbf{x}, \mathbf{h}_{i}^{(0)}\right), \quad \alpha_{i} = e^{-\lambda_{i} \Delta t}
\end{equation}
where $\Delta t = 1$ is absorbed into the network weights by default. This mechanism enables neurons to respond to external inputs while maintaining internal coupling, explicitly capturing the competition between exponential state decay and input-driven updates. The internal function $\phi_{i}$ adopts a linear form
\begin{equation}
\phi_{i}(\mathbf{x}, \mathbf{h}) = \sigma \left( \sum_{j=1}^{d} W_{i j}^{(\mathrm{recur})} h_{j} + \sum_{k=1}^{2} W_{i k}^{(\mathrm{in})} x_{k} + b_{i} \right)
\end{equation}
where $W^{(\mathrm{in})} \in \mathbb{R}^{d \times 2}$ projects spatial coordinates to neurons, and $W^{(\mathrm{recur})} \in \mathbb{R}^{d \times d}$ is a learnable recurrent matrix controlling state propagation. The input vector $\mathbf{x} = (x, y)^{T}$ undergoes linear transformation by $W^{(\mathrm{in})}$, combines with neuron states, and passes through nonlinear activation to produce final responses.

Gating mechanisms are employed at all layers for state propagation and memory control, as illustrated in Fig.~\ref{F1} (b), following a structure similar to conventional Long Short-Term Memory (LSTM) networks. At each layer, gating vectors $\mathbf{i}, \mathbf{f}, \mathbf{o} \in \mathbb{R}^{d}$ represent the input gate, forget gate, and output gate, updated through the equations
\begin{equation}
\begin{aligned}
\mathbf{i} & = \sigma\left(W_{i} \mathbf{z} + b_{i}\right) \\
\mathbf{f} & = \sigma\left(W_{f} \mathbf{z} + b_{f}\right) \\
\mathbf{o} & = \sigma\left(W_{o} \mathbf{z} + b_{o}\right) \\
\tilde{\mathbf{c}} & = \tanh\left(W_{c} \mathbf{z} + b_{c}\right)
\end{aligned}
\end{equation}
where $\mathbf{z} = \left[ \mathbf{x}; \mathbf{h}_{0} \right] \in \mathbb{R}^{d+2}$ concatenates the current input and previous hidden state $\mathbf{h}_0$. $W_{i,f,o,c}$ are weight matrices controlling input writing, memory forgetting, hidden output, and candidate memory generation, respectively; $b_{i,f,o,c}$ are bias vectors. The state updates follow
\begin{equation}
\begin{aligned}
\mathbf{c} & = \mathbf{f} \odot \mathbf{c}_{0} + \mathbf{i} \odot \tilde{\mathbf{c}} \\
\mathbf{h} & = \mathbf{o} \odot \tanh(\mathbf{c})
\end{aligned}
\end{equation}
with $\odot$ denoting element-wise multiplication, $\mathbf{c}_0$ the previous cell state, $\mathbf{c}$ the updated cell state, and $\mathbf{h}$ the current hidden state. This gating structure enables selective memory, forgetting, and output of local spatial perturbations, thereby enhancing local generalization capabilities in PEB process heat transfer modeling.

We aggregate the typed residual channels on the same computation graph using the loss definition from Eq.~(S151) of the Supplementary Material (SM).

To align training signals with the operator itself, residual channels are measured in the same inner-product structure that defines the weak energy, producing an operator-aligned objective, so we have
\begin{equation}
\label{eq:main-loss}
\begin{aligned}
    \mathcal{L}(T_\theta) &= w_1^2 \int_\Omega \langle K^{-1} \mathcal{R}_{\text{twist}}, \mathcal{R}_{\text{twist}} \rangle \mathrm{d}V_g + w_2^2 \int_\Omega h_\Omega^2 |\mathcal{R}_{\text{div}}|^2 \mathrm{d}V_g \\&+ w_3^2 \int_\Omega \langle K \mathcal{R}_{\text{cl}}, \mathcal{R}_{\text{cl}} \rangle \mathrm{d}V_g+ w_4^2 \int_{\partial \Omega} h_\Gamma |\mathcal{R}_{\text{port}}|^2 \mathrm{d}S_g ,
\end{aligned}
\end{equation}
where \(\mathcal{R}_{\mathrm{twist}}=K\nabla e,\; \mathcal{R}_{\mathrm{div}}=\nabla\!\cdot(K\nabla e),\; \mathcal{R}_{\mathrm{port}}=n\!\cdot\!K\nabla e+h\,\gamma_0 e\) as defined in Eqs.~(S146)-(S151) of the Supplementary Material (SM). Invoking the two-sided equivalence from the Eqs.~(S148), (S161) and (S162) of the Supplementary Material (SM) aligns the training objective with the energy error
\begin{equation}
    c_{\mathrm{low}}\;\|e\|_{E}^{2} \;\le\; \mathcal{L}(T_\theta) \;\le\; c_{\mathrm{up}}\;\|e\|_{E}^{2} \quad (\text{osc}=0),
\end{equation}
with constants depending only on \((k_0,\|K\|_\infty,h_0,\|h\|_\infty)\) and geometric regularity of \(\Omega\). The rigorous relations among these three components and all constant dependencies are detailed in the Sec. II of the Supplementary Material (SM). This two-sided control certifies operator-level alignment of the objective: interior, constitutive, and port channels decay in lockstep under the same energy norm, eliminating competition between PDE and trace terms.

The complete set of learnable parameters is defined as
\begin{equation}
\theta = \left\{ W^{(\mathrm{in})}, b^{(\mathrm{in})}, W_{i}, U_{i}, b_{i}, W_{f}, U_{f}, b_{f}, W_{o}, U_{o}, b_{o}, W_{c}, U_{c}, b_{c}, W^{(\mathrm{recur})}, \lambda, W^{(\mathrm{out})}, b^{(\mathrm{out})} \right\}
\end{equation}
which the parameterization remains minimal and acts only through operator-aligned gates and readouts.

In implementation, \texttt{model.parameters()} automatically collects this set for backpropagation. 

The training objective minimizes the sum of the predicted temperature field and PDE residuals using the Adam optimizer. At step $t$, first- and second-moment estimates update parameters through

\begin{equation}
\begin{aligned}
\mathbf{m}_{t} & = \beta_{1} \mathbf{m}_{t-1} + (1 - \beta_{1}) \nabla_{\theta} \mathcal{L}_{t} \\
\mathbf{v}_{t} & = \beta_{2} \mathbf{v}_{t-1} + (1 - \beta_{2}) (\nabla_{\theta} \mathcal{L}_{t})^{2} \\
\hat{\mathbf{m}}_{t} & = \frac{\mathbf{m}_{t}}{1 - \beta_{1}^{t}}, \quad \hat{\mathbf{v}}_{t} = \frac{\mathbf{v}_{t}}{1 - \beta_{2}^{t}} \\
\boldsymbol{\theta}_{t+1} & = \boldsymbol{\theta}_{t} - \eta \cdot \frac{\hat{\mathbf{m}}_{t}}{\sqrt{\hat{\mathbf{v}}_{t}} + \epsilon}
\end{aligned}
\end{equation}
with default values $\beta_{1} = 0.9$, $\beta_{2} = 0.999$, $\epsilon = 10^{-8}$, and learning rate $\eta$. Gradients $\nabla_{\theta} \mathcal{L}{t}$ are computed for the total loss $\mathcal{L}{t}$ at step $t$. We employ a fixed learning rate and train the model for 50,000 iterations to ensure both data fitting at observation points and adherence to physical constraints under non-explicit boundary conditions. This iteration count exceeds the plateau point of operator residual-energy trajectories across all backbones, matches the normalization in Tab.~\ref{tab:model-snapshot} (time/epoch $=$ total$/50{,}000$), and ensures a uniform computational budget for cross-model RMSE comparisons.

The overall architecture implements "coordinates $\to$ fields $\to$ operator/trace" as a single computational chain, where the mixed backbone provides provable conditioning control within the network without altering the geometry–port–operator pipeline or the symbol system. The Sec. III of the Supplementary Material (SM) records three theorems corresponding to this section: preconditioning equivalence, non-expansiveness with a second-order upper bound, and affine covariance.

In summary, operator alignment appears at three levels: (i) \textit{typed representation} $\Phi_\theta : \Omega \to H^1(\Omega)$; (ii) \textit{aligned dynamics} via input-conditioned diagonal preconditioning and spectral small-gain; and (iii) \textit{aligned objective} through energy-equivalent weak residuals that bind interior and port channels.

\section{Geometric–Port Formulation for PEB Heat-Flux and Objective Construction}
Let us focus on the heat-flux pathway during the post-exposure bake (PEB) stage of lithographic processing, where the wafer behaves as a two-dimensional conductive plate governed by an isotropic law $K(x)=k,\mathrm{Id}$ with Robin (convection) boundary conditions at ambient temperature $T_\infty$. We consider a circular domain with radius $R=150~\mathrm{mm}$ and employ the following physical constants in Eqs.~\eqref{9}-\eqref{14}: $k=159~\mathrm{W,m^{-1}K^{-1}}$, $Q=2000~\mathrm{W,m^{-3}}$, $h=50~\mathrm{W,m^{-2}K^{-1}}$, and $T_\infty=800~\mathrm{K}$.
The mathematical framework operates on an oriented Riemannian manifold with boundary $(\Omega,g)$ and outward unit normal $n$, where the Hodge star induced by $g$ is denoted by $*$ (with the subscript $g$ omitted hereafter). Temperature $T\in H^1(\Omega)$ is treated as a $0$-form, flux covector $\iota=q^\flat\in\Lambda^1(\Omega)$ as a $1$-form, and the $K$-twist $\omega:=K^{-1}\iota\in\Lambda^1(\Omega)$. We utilize the exterior derivative $d$, codifferential $\delta:=-*d*$, trace operators $\gamma_0:H^1(\Omega)\to H^{1/2}(\partial\Omega)$ and $\gamma_n:H(\mathrm{div};\Omega)\to H^{-1/2}(\partial\Omega)$, and set $\lambda:=\gamma_0T$. With the bundle connection $A\equiv0$ ensuring $d_A=d$ and $F_A=0$, we collect state and port variables into
\begin{equation}
    U:=(\omega,\;\iota,\;T,\;\lambda),
\end{equation}
where $T\in H^1(\Omega)$ is the temperature potential (a $0$–form), $\omega\in L^2\Lambda^1(\Omega)$ is the $K$–twist $1$–form, $\iota\in H(\mathrm{div};\Omega)$ is the heat–flux covector with well–defined normal trace $\gamma_n\iota\in H^{-1/2}(\partial\Omega)$, and $\lambda:=\gamma_0T\in H^{1/2}(\partial\Omega)$ is the Dirichlet trace. The ordering $(\omega, \iota, T, \lambda)$ aligns constitutive, balance, and port channels with the operator-aligned residual heads used in the objective. We denote by $d$ the exterior derivative, by $\delta:=-*\,d\,*$ the codifferential, and by $\nabla T:=dT$ the metric gradient $1$–form. The material tensor $K(x)$ is symmetric positive definite with lower bound $k_0>0$, and the port impedance $h(x)\ge h_0>0$. Using this notation, we assemble the constitutive-operator system in block form
\begin{equation}\label{9}
    \mathbb{A}\,U=\begin{bmatrix}
d & 0 & 0 & 0\\
0 & \delta & 0 & 0\\
I & 0 & \nabla & 0\\
0 & \gamma_n & 0 & -h\\
0 & 0 & -\gamma_0 & I
\end{bmatrix}
\!\begin{bmatrix}\omega\\ \iota\\ T\\ \lambda\end{bmatrix}
=
\begin{bmatrix}
0\\ Q\,\mathrm{vol}_g\\ 0\\ h\,T_\infty\\ 0
\end{bmatrix}.
\end{equation}
Componentwise, this system expands to
\begin{equation}\label{10}
    d\,\omega=0\ \text{in }\Omega,\quad
\delta\,\iota=Q\,\mathrm{vol}_g\ \text{in }\Omega,\quad
\iota(n)=h(\lambda-T_\infty)\ \text{on }\partial\Omega,\quad
\omega+\nabla T=0\ \text{in }\Omega,\quad
\lambda=\gamma_0T.
\end{equation}
Which composes the de Rham complex with trace and Robin impedance so that each row coincides with one residual channel in the operator-aligned loss Eqs.\eqref{12}–\eqref{14}.

Using the same notation, we define the $K$–twist residual vector $\mathcal{R}(U):=\mathbb{A}U-\big(0,\,Q\,\mathrm{vol}_g,\,0,\,hT_\infty,\,0\big)^{\!\top}$ with entries
\begin{equation}\label{11}
    \mathcal{R}
=
\begin{bmatrix}
\mathcal{R}_{\mathrm{clo}}\\ \mathcal{R}_{\mathrm{div}}\\ \mathcal{R}_{\mathrm{twist}}\\ \mathcal{R}_{\mathrm{port}}\\ \lambda-\gamma_0T
\end{bmatrix}
=
\begin{bmatrix}
d\,\omega\\[2pt]
\delta\,\iota - Q\,\mathrm{vol}_g\\[2pt]
\omega+\nabla T\\[2pt]
\iota(n)-h(\lambda-T_\infty)\\[2pt]
0
\end{bmatrix},
\end{equation}
measuring all interior terms by the Hodge inner product induced by $g$. This approach maintains consistent units (energy-density integrated over volume or boundary) and aligns operator blocks with the weak formulation used subsequently, with detailed proofs and constants provided in the Sec. IV of the Supplementary Material (SM).

We define the PDE and boundary losses as
\begin{equation}\label{12}
    \mathcal{L}_{\mathrm{PDE}}
= w_1\!\int_{\Omega}\! \mathcal{R}_{\mathrm{twist}}\wedge * \mathcal{R}_{\mathrm{twist}}
+ w_2\!\int_{\Omega}\! \mathcal{R}_{\mathrm{div}}\wedge * \mathcal{R}_{\mathrm{div}}
+ w_3\!\int_{\Omega}\! \mathcal{R}_{\mathrm{clo}}\wedge * \mathcal{R}_{\mathrm{clo}},
\end{equation}
\begin{equation}\label{13}
    \mathcal{L}_{\mathrm{BC}}
= w_4\!\int_{\partial\Omega}\! |\mathcal{R}_{\mathrm{port}}|^2\,\mathrm{d}s,
\end{equation}
and take the objective as the sum
\begin{equation}\label{14}
\mathcal{L}=\mathcal{L}_{\mathrm{PDE}}+\mathcal{L}_{\mathrm{BC}}.
\end{equation}
This alignment eliminates competition between PDE and trace terms and synchronizes their decay during training. Weights $(w_1,\dots,w_4)$ follow non-dimensional scaling described below, where coercivity of the induced energy norm and the two-sided equivalence between $\mathcal{L}$ and the weak energy error depend only on $(k_0,|K|\infty,h_0,|h|_\infty)$ and geometric regularity of $\Omega$, with complete statements and proofs in the Sec. V of the Supplementary Material (SM). 

Non-dimensionalization employs scaled coordinates with \(x' = x/L\), \(T' = T/(Q_0 L^2/k_0)\), \(K' = K/k_0\), \(h' = hL/k_0\), \(Q' = Q/Q_0\). After rewriting all integrals in the \((x',\cdot)\) frame, we set
\begin{equation}
w_1 \simeq 1, \quad w_2 \simeq \|K'\|_\infty^{-1}, \quad w_3 \simeq \varepsilon_{\text{clo}}, \quad w_4 \simeq 1,
\end{equation}
using \(\varepsilon_{\text{clo}}\) to suppress the deviation of \(\omega\) relative to \(-\nabla T\), matching the port penalty with the volumetric terms in order, and concentrating anisotropy strength in \(\|K'\|_\infty\). The two–sided constants depend only on \((k_0,\|K\|_\infty,h_0,\|h\|_\infty)\) and domain geometry; we place the full proof in the Sec. VI of the Supplementary Material (SM).

To establish verifiable local energy relations, we define the absolute error map and pointwise alignment of the three residuals. Letting \(T^*\) denote the true solution, \(e:=T-T^*\) the error, and the energy norm \(\|e\|_E^2=\int_\Omega \langle K\nabla e,\nabla e\rangle+\int_{\partial\Omega} h(\gamma_0 e)^2\). Take a smooth partition of unity \(\{\phi_i\}\) covering \(\Omega\), each \(\phi_i\) supported in a micro–patch \(\omega_i\) of diameter \(\simeq h_i\), and set \(\Gamma_i:=\partial\Omega\cap\omega_i\). To make alignment verifiable at patch scale, define operator-aligned local indicators that combine the three residual channels with geometry-consistent scaling, so we define the local indicator
\begin{equation}
\eta_i^2 := \alpha_1 \int_{\omega_i} \|\mathcal{R}_{\text{twist}}\|^2 + \alpha_2 h_i^2 \int_{\omega_i} \|\mathcal{R}_{\text{div}}\|^2 + \alpha_3 h_i \int_{\Gamma_i} |\mathcal{R}_{\text{port}}|^2,
\end{equation}
where \(\alpha_{1,2,3}\) depend only on \((k_0,\|K\|_\infty,h_0,\|h\|_\infty)\) and shape regularity. In the Sec. VII of the Supplementary Material (SM) we prove the two–sided estimate
\begin{equation}
c_1 \sum_i \eta_i^2 \leq \|e\|_E^2 \leq c_2 \sum_i \eta_i^2,
\end{equation}
and we provide the “local reliability/effectivity’’ version
\begin{equation}
c_1' \eta_i^2 \leq \int_{\omega_i} \langle K \nabla e, \nabla e \rangle + \int_{\Gamma_i} h(\gamma_0 e)^2 \leq c_2' \eta_i^2.
\end{equation}
These bounds establish a one-to-one correspondence between spatial distribution of error-energy density and local densities of $|\mathcal{R}_{\text{twist}}|^2$, $h_i^2|\mathcal{R}_{\text{div}}|^2$, and $h_i|\mathcal{R}_{\text{port}}|^2$. Within any $\omega_i$, the peaks and valleys of error maps $|e|$ and $|\nabla e|$ align with those of $\eta_i$ within constant factors and the local smoothing-kernel scale $h_i$, providing precise meaning to "pointwise alignment." Physically, the volumetric equation residual $\mathcal{R}_{\text{div}}$ releases $h_i$-scale energy inside $\omega_i$ via the Green kernel, the port residual $\mathcal{R}_{\text{port}}$ conducts inward along $\Gamma_i$ forming boundary-layer error ridges, and the constitutive residual $\mathcal{R}_{\text{twist}}$ enforces consistency between $\nabla T$ and flux, thereby locking error peaks to geometric locations with strongest gradient mismatch. The Sec. VIII of the Supplementary Material (SM) provides precise lemmas localizing the $H^{-1}/H^{-1/2}$ terms to $\omega_i$ through partition-of-unity lifting, Bogovski\u{\i} operator, and stable extension, while explaining why smooth network predictors admit negligible interior jump terms that yield the scalings $h_i^2$ and $h_i$.

For quantitative validation, we employ a MATLAB finite-element solution as a noise-free reference field, computing two key metrics at identical sample points: the root mean square error\footnote{For convenience, in Sec.~IV and thereafter we denote $\mathrm{RMSE}_T$ by RMSE.
}
\begin{equation}
\mathrm{RMSE}_T = \left( \frac{1}{N_{\text{ref}}} \sum_{p=1}^{N_{\text{ref}}} |T(x_p) - T_{\text{FEM}}(x_p)|^2 \right)^{1/2},
\end{equation}
and the relative boundary error\footnote{The algorithm computes the operator-aligned boundary metric $\mathrm{RelErr}_{\partial\Omega}$ online and uses it as an internal alignment-control signal that strengthens boundary training. Because the objective already penalizes the Robin port in the same $h$-weighted boundary norm and enjoys two-sided energy equivalence with the weak energy (cf. Eqs. \eqref{12}–\eqref{14}, the reported port-residual curves quantify boundary fidelity up to geometry-dependent constants. A separate table of $\mathrm{RelErr}_{\partial\Omega}$ would duplicate that information and introduce scale bias due to boundary-measure and nondimensionalization choices; the study therefore treats $\mathrm{RelErr}_{\partial\Omega}$ as an internal safeguard rather than a stand-alone reported metric.
}
\begin{equation}
\mathrm{RelErr}_{\partial\Omega}
=
\frac{\big(\int_{\partial\Omega}\!|\gamma_n \iota + h(\gamma_0 T - T_\infty)|^2\,\mathrm{d}s\big)^{1/2}}
     {\big(\int_{\partial\Omega}\!|h(\gamma_0 T - T_\infty)|^2\,\mathrm{d}s\big)^{1/2}}.
\end{equation}
These metrics provide complementary insights, with $\mathrm{RMSE}_T$ quantifying global deviation and $\mathrm{RelErr}_{\partial\Omega}$ measuring relative mismatch in port conditions, while absolute-error heat maps reveal spatial error structure. To establish correlation between local residuals and global errors, we rank blockwise indicators ${\eta_i}$ in descending order and verify that the top $k$ peaks in absolute-error heat maps reside within the corresponding micro-patch set $\bigcup_{i\in I_k}\omega_i$, with peak intensities scaling proportionally to $\eta_i$ up to multiplicative constants.
The connection between domain-wide energy density and local residual densities is rigorously established through local two-sided bounds. The Supplementary Material (SM) prove the reliability/effectivity bound (Eq. (S187))
\begin{equation}
c_1'\eta_i^2 \le\int_{\omega_i}\langle K\nabla e,\nabla e\rangle\mathrm{d}x
+ \int_{\Gamma_i}h(\gamma_0 e)^2\mathrm{d}s
\le c_2'\eta_i^2 .
\end{equation}
To capture boundary-layer geometry at the \(h_i\) scale, the Supplementary Material (SM) defines smoothed density maps (Equtaions~(S195)–(S196) of the Supplementary Material (SM))
and proves the pointwise two-sided control on the shrunken core \(\omega_i^\circ\) (Eq.~(S197) of the Supplementary Material (SM))
\begin{equation}
    \begin{aligned}
&E_i(x) := S_i\xi(x) + S^{\Gamma}_i\zeta(x),\\
&\mathcal{R}_i(x) := S_i\!\langle K^{-1}\mathcal{R}_{\mathrm{twist}},\mathcal{R}_{\mathrm{twist}}\rangle
         + S_i\!\big(h_i^{2}|\mathcal{R}_{\mathrm{div}}|^{2}\big)
         + S^{\Gamma}_i\!\big(h_i|\mathcal{R}_{\mathrm{port}}|^{2}\big),\\
\hspace{4.2cm} &c_{\min}\,E_i(x) \;\le\; \mathcal{R}_i(x) \;\le\; c_{\max}\,E_i(x)\qquad (x\in\omega_i^\circ).
\end{aligned}
\end{equation}
These inequalities collectively quantify the strict alignment of boundary-layer ridges with port-dominant regions, where error ridges align precisely with regions on $\Gamma_i$ where port terms contribute most significantly. All proofs and constant dependencies, which rely only on $(k_0,|K|\infty,h_0,|h|\infty)$ and shape regularity, are provided in the Sec. IX of the Supplementary Material (SM).

In summary, the geometric–port system, residual definitions, weights, and local indicators all act in operator-aligned coordinates, so training signals and error diagnostics remain coherent from global energy down to patch-scale structure.
\section{Quantitative Validation under Geometry–Port Constraints}
\begin{figure*}[!ht]
	\centering
	\includegraphics[width=0.8\textwidth]{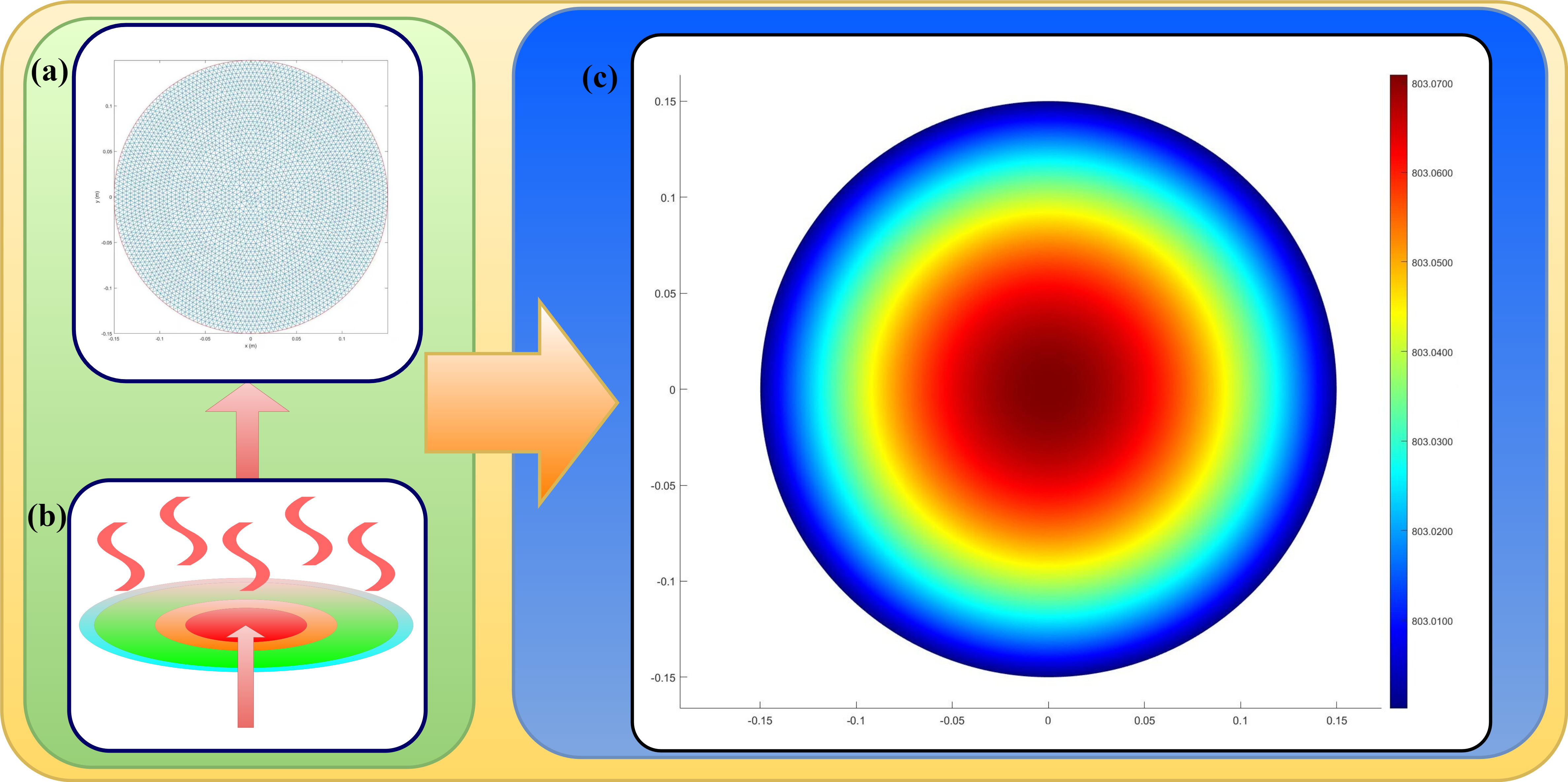}
	\caption{Temperature distribution obtained by numerically simulating the PEB process heat transfer problem using the finite element method on the MATLAB platform. This result serves as a reference benchmark for subsequent computation of RMSE and absolute error distributions for different model predictions. The figure illustrates the spatial distribution of the temperature field within the computational domain, reflecting the effects of finite element discretization and boundary conditions on PEB process thermal behavior. \textbf{(a)} We depict the post-exposure bake (PEB) setup: a circular wafer domain $\Omega$ (radius $R$) with internal heat generation $Q$ and a convective (Robin) boundary characterized by coefficient $h$ and ambient temperature $T_\infty$; arrows indicate heat flux and boundary exchange. \textbf{(b)} We show the finite-element (FEM) mesh that provides the reference solution, i.e., a conforming triangular tessellation of the circular domain with characteristic element size $H_{\max}$, respecting the geometry and boundary ports. \textbf{(c)} We present representative outcomes under this setup: the steady-state temperature field $T(x,y)$ and the absolute-error map $\lvert T_{\text{model}}(x,y)-T_{\text{FEM}}(x,y)\rvert$, which we use for visual diagnostics and RMSE evaluation.}
	\label{F2}
\end{figure*}
\begin{table*}[!ht]
  \centering
  \caption{The reported learning rate for each method is the manuscript optimum obtained from the unified sweep-and-select protocol. Snapshot comparison across representative learning rates (identical hardware/schedule). We report total wall-clock time, mean time per epoch (s/epoch $=$ total$/50{,}000$), and RMSE; values use scientific notation with two decimals.}
  \label{tab:model-snapshot}
  \begin{tabular*}{\textwidth}{@{\extracolsep{\fill}} l c r r r}
    \toprule
    Model & $LR$ & training time (s) & time/epoch (s) & RMSE \\
    \midrule
    PINN & 1e-4 & 4.50e+04 & 9.01e-01 & 1.13e-04 \\
    LNN-PINN & 7e-4 & 5.12e+04 & 1.02e+00 & 8.23e-05 \\
    LSTM-PINN & 1e-3 & 5.97e+03 & 1.19e-01 & 8.96e-05 \\
    LSTM-LNN-PINN & 8e-4 & 6.36e+03 & 1.27e-01 & 3.83e-05 \\
    \bottomrule
  \end{tabular*}
\end{table*}
\begin{table*}[!ht]
  \centering
  \caption{Learning-rate sweep for the operator-aligned PINN surrogate for PEB heat-flux on silicon wafers: wall-clock training time, per-epoch time, and accuracy metrics (RMSE, MAE, MSE).}
  \label{tab:pinn-runtime-error-vs-lr}
  \begin{tabular*}{\textwidth}{@{\extracolsep{\fill}} l r r r r r}
    \toprule
    $LR$ & training time (s) & time/epoch (s) & RMSE & MAE & MSE \\
    \midrule
    1e-4 & 4.50e+04 & 9.01e-01 & 1.13e-04 & 9.38e-05 & 1.27e-08 \\
    2e-4 & 4.50e+04 & 9.01e-01 & 1.98e-04 & 1.87e-04 & 3.90e-08 \\
    3e-4 & 4.50e+04 & 9.00e-01 & 2.66e-04 & 2.21e-04 & 7.09e-08 \\
    4e-4 & 4.50e+04 & 9.01e-01 & 3.34e-04 & 3.27e-04 & 1.11e-07 \\
    5e-4 & 4.50e+04 & 9.00e-01 & 9.24e-04 & 8.84e-04 & 8.54e-07 \\
    6e-4 & 4.50e+04 & 9.00e-01 & 4.56e-04 & 4.08e-04 & 2.08e-07 \\
    7e-4 & 4.50e+04 & 9.00e-01 & 2.45e-04 & 2.32e-04 & 6.02e-08 \\
    8e-4 & 4.50e+04 & 9.00e-01 & 1.24e-03 & 1.21e-03 & 1.53e-06 \\
    9e-4 & 4.50e+04 & 9.00e-01 & 1.13e-04 & 1.07e-04 & 1.29e-08 \\
    1e-3 & 4.50e+04 & 9.00e-01 & 6.16e-04 & 6.14e-04 & 3.80e-07 \\
    \bottomrule
  \end{tabular*}
\end{table*}
\begin{table*}[!ht]
  \centering
  \caption{Learning-rate sweep for the operator-aligned LNN\textendash PINN surrogate for PEB heat-flux on silicon wafers: wall-clock training time, per-epoch time, and accuracy metrics (RMSE, MAE, MSE).}
  \label{tab:lnn-pinn-runtime-error-vs-lr}
  \begin{tabular*}{\textwidth}{@{\extracolsep{\fill}} l r r r r r}
    \toprule
    $LR$ & training time (s) & time/epoch (s) & RMSE & MAE & MSE \\
    \midrule
    1e-4 & 5.13e+04 & 1.03e+00 & 2.52e-04 & 2.45e-04 & 6.37e-08 \\
    2e-4 & 5.13e+04 & 1.03e+00 & 3.90e-04 & 3.83e-04 & 1.52e-07 \\
    3e-4 & 5.13e+04 & 1.03e+00 & 1.38e-03 & 1.34e-03 & 1.91e-06 \\
    4e-4 & 5.13e+04 & 1.03e+00 & 7.78e-04 & 7.16e-04 & 6.05e-07 \\
    5e-4 & 5.12e+04 & 1.02e+00 & 5.14e-04 & 5.09e-04 & 2.65e-07 \\
    6e-4 & 5.08e+03 & 1.02e-01 & 3.96e-04 & 3.94e-04 & 1.56e-07 \\
    7e-4 & 5.12e+04 & 1.02e+00 & 8.23e-05 & 6.84e-05 & 6.80e-09 \\
    8e-4 & 5.12e+04 & 1.02e+00 & 2.65e-04 & 2.17e-04 & 7.00e-08 \\
    9e-4 & 5.12e+04 & 1.02e+00 & 6.98e-04 & 6.93e-04 & 4.88e-07 \\
    1e-3 & 5.14e+04 & 1.03e+00 & 2.03e-04 & 2.00e-04 & 4.13e-08 \\
    \bottomrule
  \end{tabular*}
\end{table*}
\begin{table*}[!ht]
  \centering
  \caption{Learning-rate sweep for the operator-aligned LSTM\textendash PINN surrogate for PEB heat-flux on silicon wafers: wall-clock training time, per-epoch time, and accuracy metrics (RMSE, MAE, MSE).}
  \label{tab:lstm-pinn-runtime-error-vs-lr}
  \begin{tabular*}{\textwidth}{@{\extracolsep{\fill}} l r r r r r}
    \toprule
    $LR$ & training time (s) & time/epoch (s) & RMSE & MAE & MSE \\
    \midrule
   1e-4 & 6.21e+03 & 1.24e-01 & 1.56e-04 & 1.52e-04 & 2.44e-08 \\
    2e-4 & 6.38e+03 & 1.28e-01 & 2.67e-04 & 2.64e-04 & 7.15e-08 \\
    3e-4 & 5.97e+03 & 1.19e-01 & 1.39e-04 & 1.34e-04 & 1.94e-08 \\
    4e-4 & 5.90e+03 & 1.18e-01 & 1.43e-04 & 1.28e-04 & 2.05e-08 \\
    5e-4 & 6.15e+03 & 1.23e-01 & 5.89e-04 & 5.86e-04 & 3.47e-07 \\
    6e-4 & 6.11e+03 & 1.22e-01 & 9.64e-05 & 9.01e-05 & 9.30e-09 \\
    7e-4 & 5.99e+03 & 1.20e-01 & 6.76e-04 & 6.74e-04 & 4.57e-07 \\
    8e-4 & 5.89e+03 & 1.18e-01 & 1.01e-03 & 1.01e-03 & 1.03e-06 \\
    9e-4 & 6.00e+03 & 1.20e-01 & 2.22e-04 & 2.19e-04 & 4.92e-08 \\
    1e-3 & 5.97e+03 & 1.19e-01 & 8.96e-05 & 8.26e-05 & 8.00e-09 \\
    \bottomrule
  \end{tabular*}
\end{table*}
\begin{table*}[!ht]
  \centering
  \caption{Learning-rate sweep for the operator-aligned LSTM\textendash LNN\textendash PINN surrogate for PEB heat-flux on silicon wafers: wall-clock training time, per-epoch time, and accuracy metrics (RMSE, MAE, MSE).}
  \label{tab:lstm-lnn-pinn-runtime-error-vs-lr}
  \begin{tabular*}{\textwidth}{@{\extracolsep{\fill}} l r r r r r}
    \toprule
    $LR$ & training time (s) & time/epoch (s) & RMSE & MAE & MSE \\
    \midrule
    1e-4 & 5.68e+03 & 1.14e-01 & 8.82e-05 & 8.10e-05 & 7.80e-09 \\
    2e-4 & 5.46e+03 & 1.09e-01 & 1.48e-04 & 1.30e-04 & 2.18e-08 \\
    3e-4 & 6.10e+03 & 1.22e-01 & 1.27e-04 & 1.22e-04 & 1.61e-08 \\
    4e-4 & 5.44e+03 & 1.09e-01 & 4.19e-04 & 4.04e-04 & 1.76e-07 \\
    5e-4 & 6.34e+03 & 1.27e-01 & 1.26e-03 & 1.17e-03 & 1.60e-06 \\
    6e-4 & 5.44e+03 & 1.09e-01 & 4.25e-04 & 4.05e-04 & 1.81e-07 \\
    7e-4 & 5.91e+03 & 1.18e-01 & 1.90e-03 & 1.87e-03 & 3.62e-06 \\
    8e-4 & 6.36e+03 & 1.27e-01 & 3.83e-05 & 3.13e-05 & 1.50e-09 \\
    9e-4 & 6.27e+03 & 1.25e-01 & 2.35e-04 & 2.08e-04 & 5.54e-08 \\
    1e-3 & 6.75e+03 & 1.35e-01 & 4.15e-04 & 3.92e-04 & 1.72e-07 \\
    \bottomrule
  \end{tabular*}
\end{table*}
\begin{figure*}[!ht]
	\centering
	\includegraphics[width=1\textwidth]{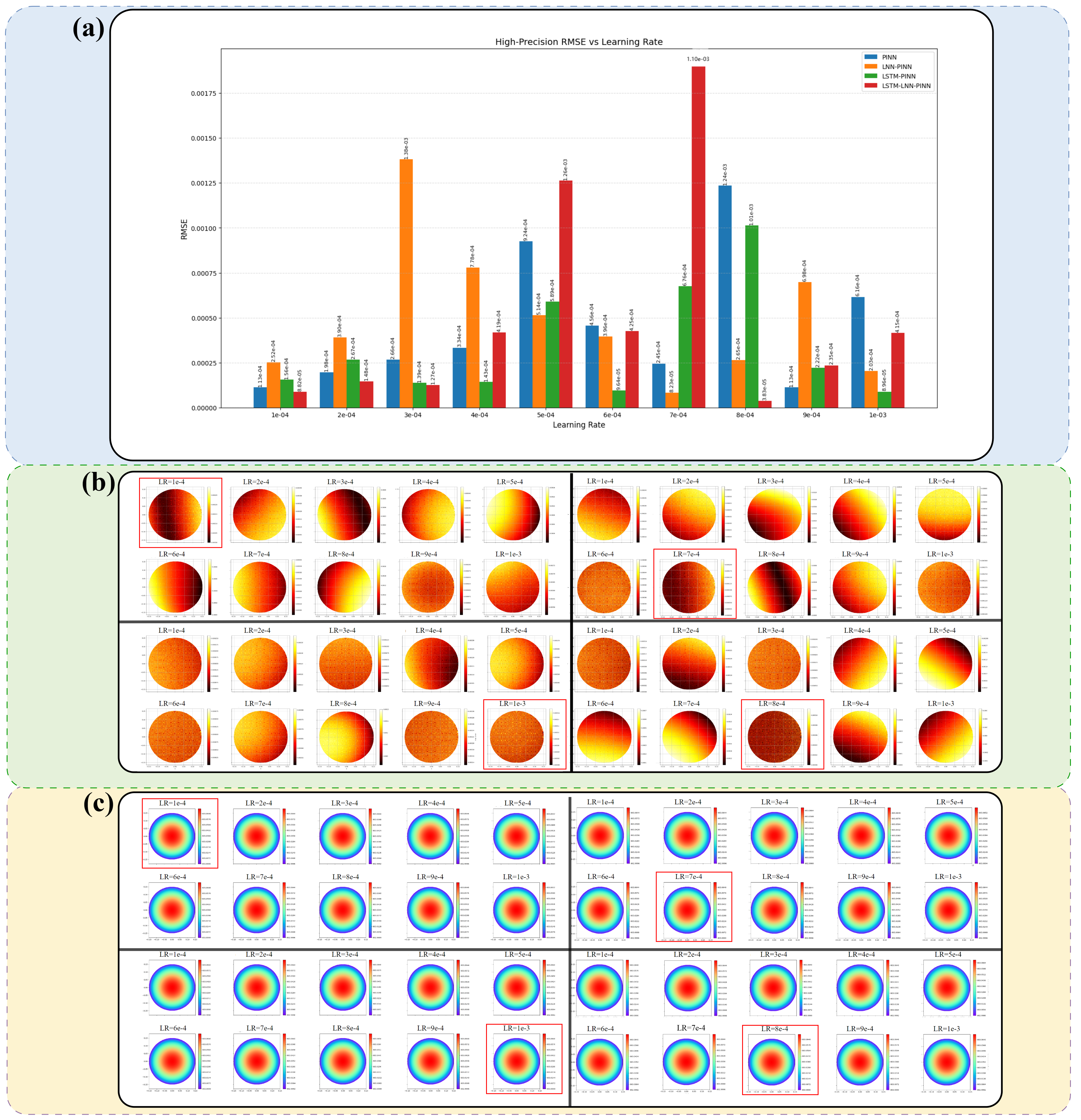}
	\caption{\textbf{(a)} RMSE versus learning rate (1e-4–1e-3; step 1e-4) for PINN, LNN-PINN, LSTM-PINN, and LSTM-LNN-PINN, highlighting robustness, sensitivity, and best-performing rates.
\textbf{(b)} Absolute-error maps relative to the FEM reference across the same rates, revealing spatial patterns and magnitudes of model errors.
\textbf{(c)} Predicted steady-state temperature fields across the same rates.}
\label{F3}
\end{figure*}
\begin{figure*}[!ht]
	\centering
	\includegraphics[width=1\textwidth]{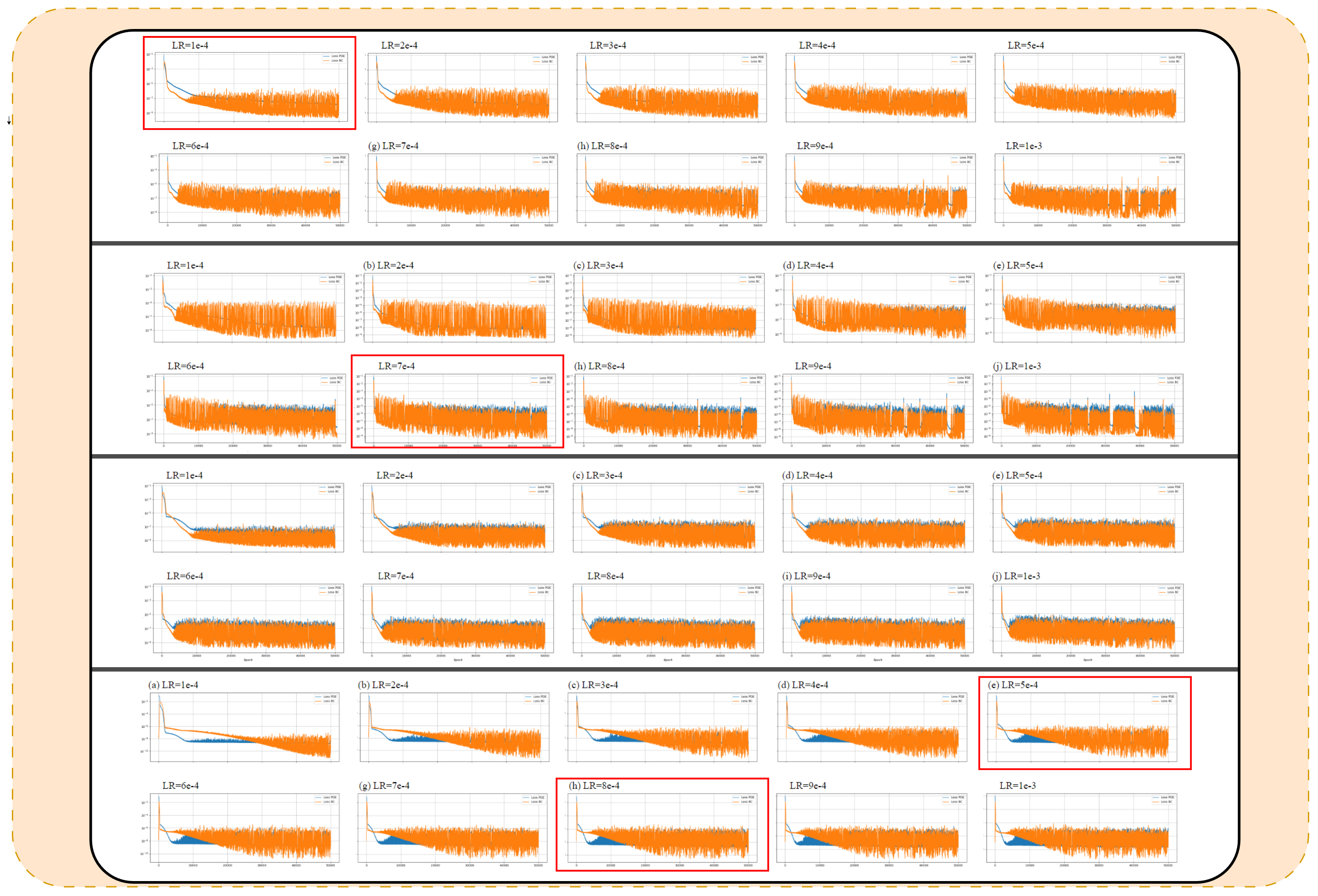}
	\caption{Training loss curves of PINN, LNN-PINN, LSTM-PINN and LSTM-LNN-PINN models under multiple learning rate settings plotted in linear coordinates. Learning rates start from 1e-4 and increase by 1e-4 up to 1e-3, with each curve representing the dynamic evolution of training loss for a specific learning rate. This figure reflects the convergence trends of the model during training and provides a visual basis for understanding the performance of the LSTM-LNN structure within the PINN framework and the effects of learning rate adjustments.}
	\label{Faa}
\end{figure*}
\begin{figure*}[!ht]
	\centering
	\includegraphics[width=1\textwidth]{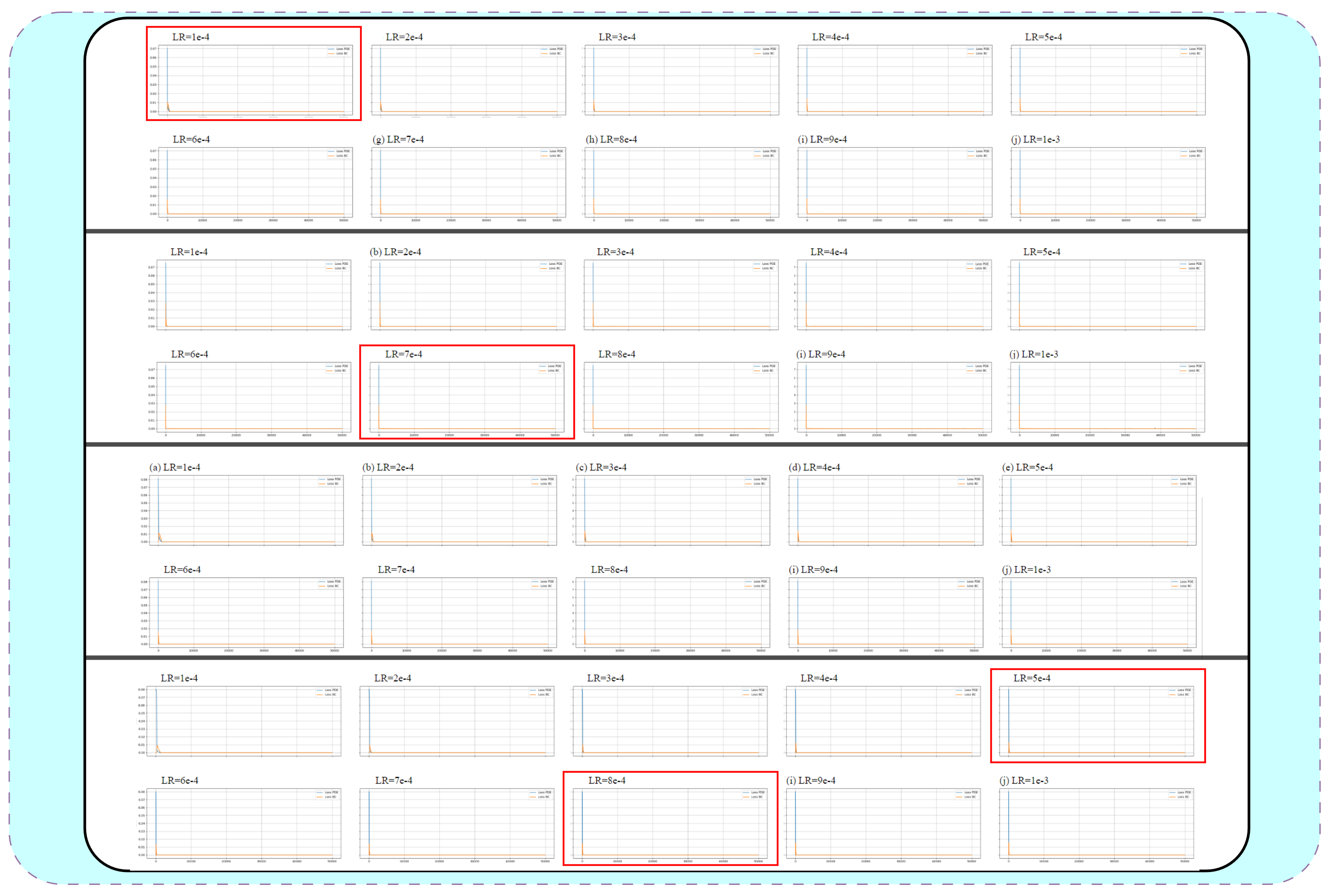}
	\caption{Training loss curves of PINN, LNN-PINN, LSTM-PINN and LSTM-LNN-PINN models under different learning rate settings plotted in logarithmic coordinates. Learning rates range from 1e-4 to 1e-3 with increments of 1e-4. The horizontal axis represents training iterations, and the vertical axis is the logarithmic scale of the loss function, highlighting magnitude changes and subtle fluctuations under different learning rates. This log-scale loss plot provides an important visual reference for evaluating the convergence performance and training stability of the LSTM-LNN-PINN model.
    }
	\label{Fbb}
\end{figure*}
We employ a finite-element method (FEM) in MATLAB to generate the reference solution shown in Fig.~\ref{F2}; discretization, solver verification and settings appear in the appendix of our companion paper ~\cite{tao2025lnn}. This FEM benchmark provides the "ground truth" for all subsequent error metrics. Following the unified selection protocol from the Figs.~\ref{F3} (b), (c), Fig.~\ref{Faa} and Fig.~\ref{Fbb}, we systematically evaluate each backbone architecture. This involves sweeping learning rates (LR) from $10^{-4}$ to $10^{-3}$ in increments of $10^{-4}$ under identical training conditions for 50,000 iterations. The optimal configuration for each model is selected based on validation RMSE, with MAE/MSE and convergence stability used as tie-breakers. This rigorous benchmarking is essential not only to find the best-performing model but also to probe each architecture's conditioning and robustness to hyperparameter choices. 

The benchmarking reveals distinct optimal learning rates and performance tiers, as consolidated in Tab.~\ref{tab:model-snapshot}. The standard PINN achieves its minimum RMSE of $1.13\times10^{-4}$ at a low $1.00\times10^{-4}$ LR, suggesting limited stability at higher rates. The LNN-PINN reaches approximately $8.23\times10^{-5}$ at $7.00\times10^{-4}$, and the LSTM-PINN (without LNN) attains about $8.96\times10^{-5}$ at $1.00\times10^{-3}$. Most notably, the hybrid {LSTM-LNN-PINN records a superior performance of $3.83\times10^{-5}$ at $8.00\times10^{-4}$}. This result is not merely an incremental improvement; its RMSE is less than half (46.5\%) of the next-best competitor (LNN-PINN), demonstrating a significant leap in predictive accuracy. The complete hyperparameter sweep and optimal configurations are documented in Figs.~\ref{F3} (b), (c), Fig.~\ref{Faa} and Fig.~\ref{Fbb}, with Tab.~\ref{tab:model-snapshot}-\ref{tab:lstm-lnn-pinn-runtime-error-vs-lr} consolidating the representative settings. 

While all four models provide qualitatively similar fits to the PEB process thermal field at the macroscopic level, as seen in the predicted temperature fields in Fig.~\ref{F3}(c), this visual similarity masks profound differences. A detailed analysis of the absolute-error maps (Fig.~\ref{F3}(b)) exposes the first critical finding. This "micro-differentiation under macro similarity" provides crucial insights for method selection in high-precision PEB process simulations. The baseline PINN exhibits significant {edge-concentrated errors} reaching $2.50\times10^{-4}$. This failure mode is particularly detrimental, as it indicates an inability to enforce boundary constraints consistently, leading to maximum error in the precise region where critical dimension control is paramount in lithography. The LNN-PINN reduces this maximum error to $2.00\times10^{-4}$ but maintains significant local variations. The LSTM-PINN further mitigates error aggregation with a maximum of $1.60\times10^{-4}$, though its uniformity remains suboptimal. In stark contrast, the {LSTM-LNN-PINN is the only architecture to achieve superior performance with errors below $1.00\times10^{-4}$ {uniformly distributed} across the entire domain}. This demonstrates a unique capacity to suppress the systematic, spatially-concentrated bias that plagues other baselines. It is not just that the average error (RMSE) is lower; it is that the {structure} of the error is fundamentally different and non-systematic, indicating a more physically consistent solution. 

This superior error profile and the suppression of boundary artifacts are fundamentally explained by the second key finding, revealed in the comparative analysis of loss convergence in Fig.~\ref{Faa} and Fig.~\ref{Fbb}. While all models exhibit similar overall loss reduction in linear scale (Fig.~\ref{Faa}), the {logarithmic scaling (Fig.~\ref{Fbb}) reveals fundamental differences in training stability} and operator-level consistency. The PINN baseline ($1.00\times10^{-4}$) converges but suffers from {late-stage fluctuations} in both PDE and boundary condition losses, indicating a persistent conflict between fitting the interior and the boundary. The LNN-PINN ($7.00\times10^{-4}$) displays {severe oscillations}, indicating clear stability limitations and ill-conditioning, likely from uncontrolled gradient dynamics. Even the LSTM-PINN ($1.00\times10^{-3}$), while better, still maintains a stable PDE loss but continues to {oscillate in its boundary conditions}, suggesting that gating alone is insufficient without the correct operator-aligned structure. 

Only the {LSTM-LNN-PINN ($8.00\times10^{-4}$) demonstrates comprehensive advantages}. Its PDE loss rapidly drops to a minimal plateau ($\sim 1.00\times10^{-6}$), its boundary condition loss converges {synchronously} without conflict, and overall fluctuations remain substantially smaller than all other architectures. This synchronous, stable convergence is direct evidence of a well-conditioned optimization problem, where the model's architecture (specifically its decoupled gating and liquid dynamics as argued above) properly aligns the computational graph with the physical operators. The network is not fighting itself; it learns the PDE interior and the Robin-port boundary conditions as a single, unified problem. These log-scale observations provide the definitive link between training dynamics and predictive accuracy, confirming that the LSTM-LNN-PINN's unique stability is the direct cause of its superior, uniform error profile. This alignment of findings, from the top-level RMSE, to the uniform error distribution, to the stable convergence, fundamentally explains its leading performance in high-fidelity PEB process thermal field reconstruction.
\section{Conclusion}
This study has established an operator-aligned LSTM-LNN-PINN framework for high-fidelity thermal reconstruction in post-exposure bake processes on silicon wafers. By unifying geometric coordinates, physical fields, and differential operators within a single computation graph through gated liquid layers, our architecture achieves fundamental advances in both training stability and predictive accuracy. The experimental validation reveals two critical breakthroughs: (1) the decoupled-gating mechanism enables exceptionally stable, operator-consistent convergence where log-scale residual energy rapidly tightens to a plateau without the late-stage oscillations that plague conventional PINN architectures; (2) the model achieves superior accuracy with an RMSE of \(3.83 \times 10^{-5}\) while maintaining domain-wide absolute errors uniformly below \(1.00 \times 10^{-4}\), thereby eliminating the edge-concentrated error patterns that undermine standard baselines in boundary-critical regions. These results collectively demonstrate that operator alignment resolves fundamental conditioning issues in physics-informed learning, delivering a robust, accurate, and computationally efficient surrogate for PEB heat-flux analysis with direct applicability to other physical domains requiring operator-consistent neural approximations.

\clearpage
\printcredits
\section*{Declaration of competing interest}
The authors declared that they have no conflicts of interest to this work. 

\section*{Acknowledgment}
This work is supported by the developing Project of Science and Technology of Jilin Province (20240402042GH). 

\section*{Data availability}
All the code for this article is available open access at a Github repository available at https://github.com/Uderwood-TZ/Operator-Consistent-Physics-Informed-Learning-for-Wafer-Thermal-Reconstruction-in-Lithography.git.

\bibliographystyle{model1-num-names}
\bibliography{cas-refs}
\end{document}